\documentclass{ws-procs9x6}

\begin{document}

\title{Sivers function in constituent quark models}

\author{S. Scopetta}
\address{
Dipartimento di Fisica, Universit\`a degli Studi
di Perugia, and INFN,\\
 sezione di Perugia, via A. Pascoli,
06100 Perugia, Italy
}

\author{A. Courtoy}
\address {
Departament de Fisica Te\`orica, Universitat de Val\`encia
and IFIC, CSIC
\\
46100 Burjassot
(Val\`encia), Spain
}

\author{F. Fratini}
\address{
Dipartimento di Fisica, Universit\`a degli Studi
di Perugia,\\
via A. Pascoli,
06100 Perugia, Italy
}

\author{V. Vento}
\address {
Departament de Fisica Te\`orica, Universitat de Val\`encia
and IFIC, CSIC
\\
46100 Burjassot
(Val\`encia), Spain,\\
TH-Division, PH Department, CERN, CH-1211 Gen\`eve 23, 
Switzerland
}

\begin{abstract}
A formalism to evaluate the Sivers function, developed 
for calculations in constituent quark models, is applied to the Isgur-Karl model. 
A non-vanishing Sivers asymmetry, with opposite signs for the $u$ and $d$ flavor, is found; 
the Burkardt sum rule is fulfilled up to ~2$\%$.
Nuclear effects in the extraction of neutron
single spin asymmetries in semi-inclusive deep inelastic scattering
off $^3$He are also evaluated. In the kinematics of JLab, 
it is found that the nuclear effects described by an 
Impulse Approximation approach are under control.
\end{abstract}


\bodymatter

\section{The Sivers function in Constituent Quark Models}

The partonic structure of transversely polarized nucleons
is still an open problem\cite{bdr}.
Semi-inclusive deep inelastic scattering (SIDIS)
is one of the proposed
processes to access the parton distributions (PDs)
of transversely polarized hadrons.
\begin{figure}{t}
 \centering
  \includegraphics[height=.17\textheight]{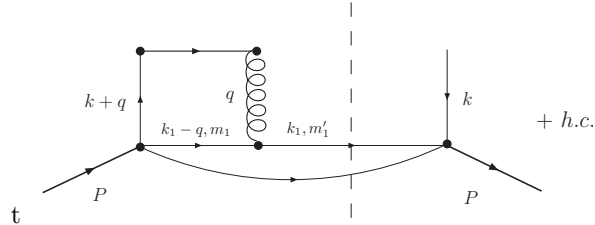}
  \caption{The contributions to the Sivers
function in the present approach.}
\end{figure}
SIDIS of unpolarized electrons off a transversely polarized target
shows "single spin asymmetries'' (SSAs) \cite{Collins},
due to two physical mechanisms,
whose contributions can be distinguished
\cite{mu-ta,ko-mu,boer}, i.e. the Collins\cite{Collins} 
and the Sivers\cite{sivers}  mechanisms.
The former is due to parton final state interactions
(FSI) in the production of a hadron
by a transversely polarized quark.
The Sivers mechanism leads to a SSA which is the product of 
the unpolarized fragmentation function with  the Sivers PD. The latter
describes the number density of unpolarized quarks
in a transversely polarized target: it is a time-reversal odd,
Transverse Momentum Dependent  (TMD) PD.
From the existence of leading-twist
Final State Interactions (FSI) \cite{brohs,brodhoy}, 
 a non-vanishing Sivers function has been explained as 
 generated by the gauge link in the definition of TMDs
\cite{coll2,adra}, whose
contribution does not vanish in the light-cone gauge,
as happens for the standard PD functions.
Recently, the first data
of SIDIS off transversely polarized targets have been published,
for the proton \cite{hermes} and the deuteron \cite{compass}.
It has been found that, while the Sivers effect
is sizable for the proton, it becomes negligible for the deuteron,
so that apparently the neutron contribution cancels the proton one,
showing a strong flavor dependence of the mechanism.
Different parameterizations of the available
SIDIS data have been published \cite{ans,coll3,Vogelsang:2005cs}, 
still with large error bars.
Since a calculation from first
principles in QCD is not yet possible, 
several model evaluations have been performed, e.g. in a quark-diquark model
\cite{brohs,coll2,bacch}; in the MIT bag model \cite{yuan}; 
in a light-cone model \cite{luma}; in a nuclear
framework, relevant to proton-proton collisions \cite{bianc}.
We here describe a Constituent Quark Model (CQM) calculation 
of the Sivers function
\cite{nostro}.
CQM calculations of PDs are based on a two steps procedure\cite{trvv}.
First, the matrix element of the
proper operator is evaluated using the wave functions of the model;
then, a low momentum scale, $\mu_0^2$, is ascribed to
the model calculation and QCD evolution is used
to evolve the observable calculated in this low energy scale to
the  scale of DIS experiments.
Such
procedure has proven successful in describing the gross features of
PDs\cite{h1}
and GPDs\cite{epj},
 by using different CQMs, e.g. the Isgur-Karl (IK) model \cite{ik}. 
 Besides the fact that it successfully reproduces the low-energy properties of the nucleon, the IK model
 contains the one-gluon-exchange (OGE)
mechanism\cite{ruju}. 
%
In the present calculation, with respect to calculations
of PDs and GPDs, the leading twist contribution to the
FSI has to be taken into account.
 The main approximations
have been: i) only the valence quark sector is investigated; ii) the leading 
twist FSI are taken into account at leading, OGE, order,
which is natural in the IK model; 
iii) the resulting interaction has been obtained
through a non-relativistic (NR) reduction of the relevant operator,
according to the philosophy of constituent quark models \cite{ruju},
leading to a potential $V_{NR}$.
The Sivers function for  a proton polarized along the $y$ axis and for the
quark of flavor ${\cal Q}$,
$f_{1T}^{\perp {\cal Q}} (x, {k_T} )$, takes the form
(cf. Fig. 1 for the labels of the momenta and helicities):
\begin{eqnarray}
&&f_{1T}^{\perp {\cal Q}} (x, {k_T} ) \nonumber\\
& &=\Im
\left \{
- i g^2
{
M^2 \over k_x
}
\int
d \vec k_1
d \vec k_3
{d^2 \vec q_T \over (2 \pi)^2}
\delta(k_3^+ - xP^+)
\delta(\vec k_{3 T} + \vec q_T - \vec k_T) {\cal M}^{\cal Q}
\right \}
\label{start2}
\end{eqnarray}
where $g$ is the strong coupling constant, $M$ the proton mass,
and
\begin{eqnarray}
{\cal M}^{u(d)} & = &
\sum_{m_1,m_1',m_3,m_3'}
\Phi_{sf,S_z=1}^{\dagger}
\left ( \vec k_3, m_3; \vec k_1, m_1;
\, \vec P - \vec k_3 - \vec k_1,  m_n  \right )
\nonumber
\\
& \times &
{ 1 \pm \tau_3(3) \over 2 }
V_{NR}(\vec k_1, \vec k_3, \vec q)
\nonumber
\\
& \times &
\Phi_{sf , S_z=-1}
\left (\vec k_3 + \vec q, m_3'; \, \vec k_1 -
\vec q, m_1';
\, \vec P - \vec k_3 - \vec k_1,  m_n  \right )~.
\label{Mu}
\end{eqnarray}

Using the spin-flavor wave function of the proton
in momentum space, 
$\Phi_{sf}$, corresponding
to a given CQM, the Sivers function,
Eq. (\ref{start2}),
can be evaluated.
From Eq.~(\ref{Mu}), one  notices that the helicity conserving
part of the global interaction
does not contribute to the Sivers function.
Besides,
in an extreme NR limit, it turns
out to be identically zero:
in our  scheme, it is precisely the interference of the small
and large components in the four-spinors
of the free quark states which leads to a non-vanishing
Sivers function. This holds even from
the component with $l = 0$ of the target wave function.
While, in other approaches\cite{yuan},  these interference terms arise due to the wave function,
they are produced here by the interaction.
The above-described formalism is now applied to the IK model. The detailed procedure and
the final expressions of the Sivers function in this model can be found
in Ref. \cite{nostro}.
\begin{figure}
\includegraphics[width=.49\textwidth]{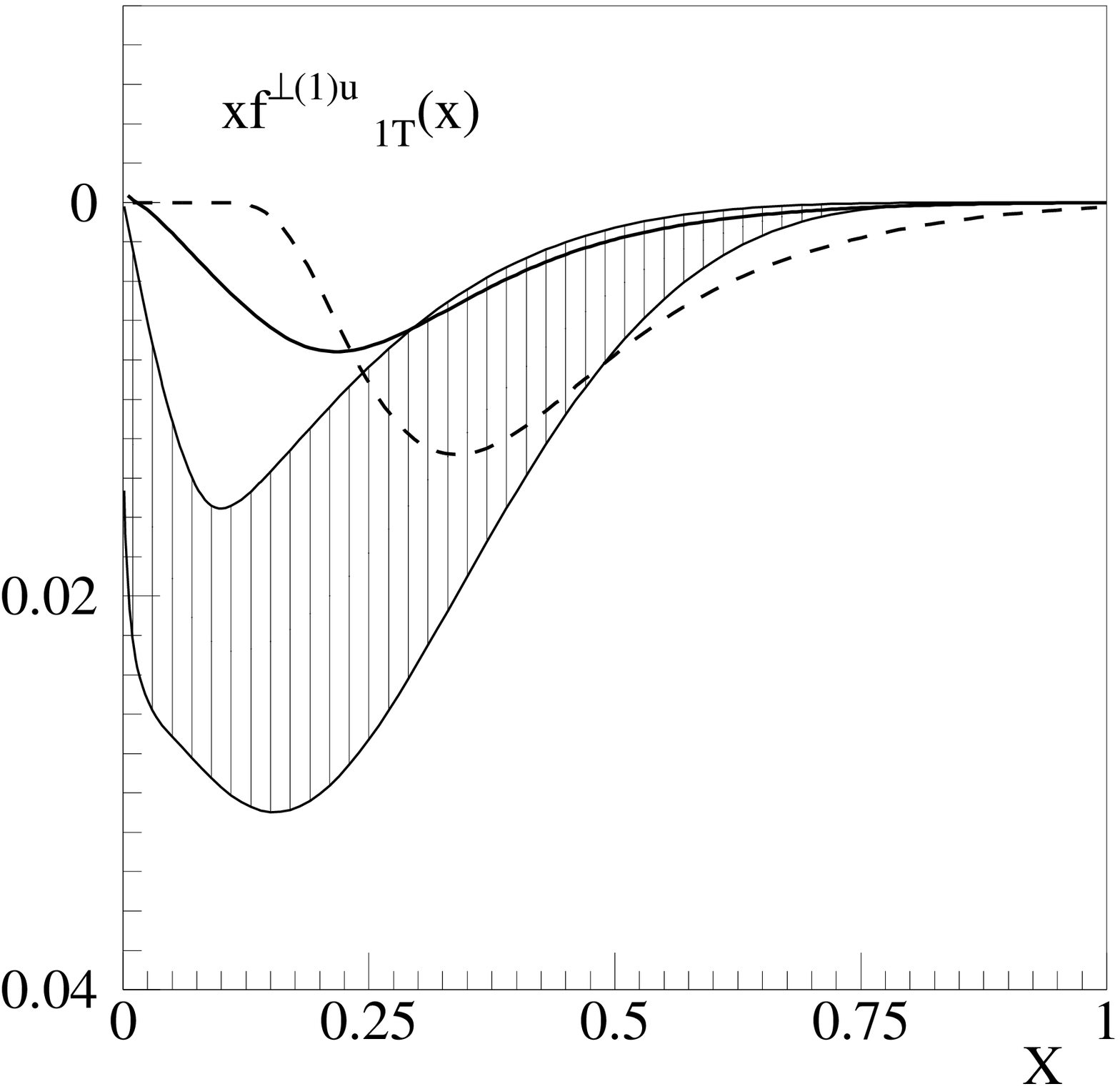}
\includegraphics[width=.49\textwidth]{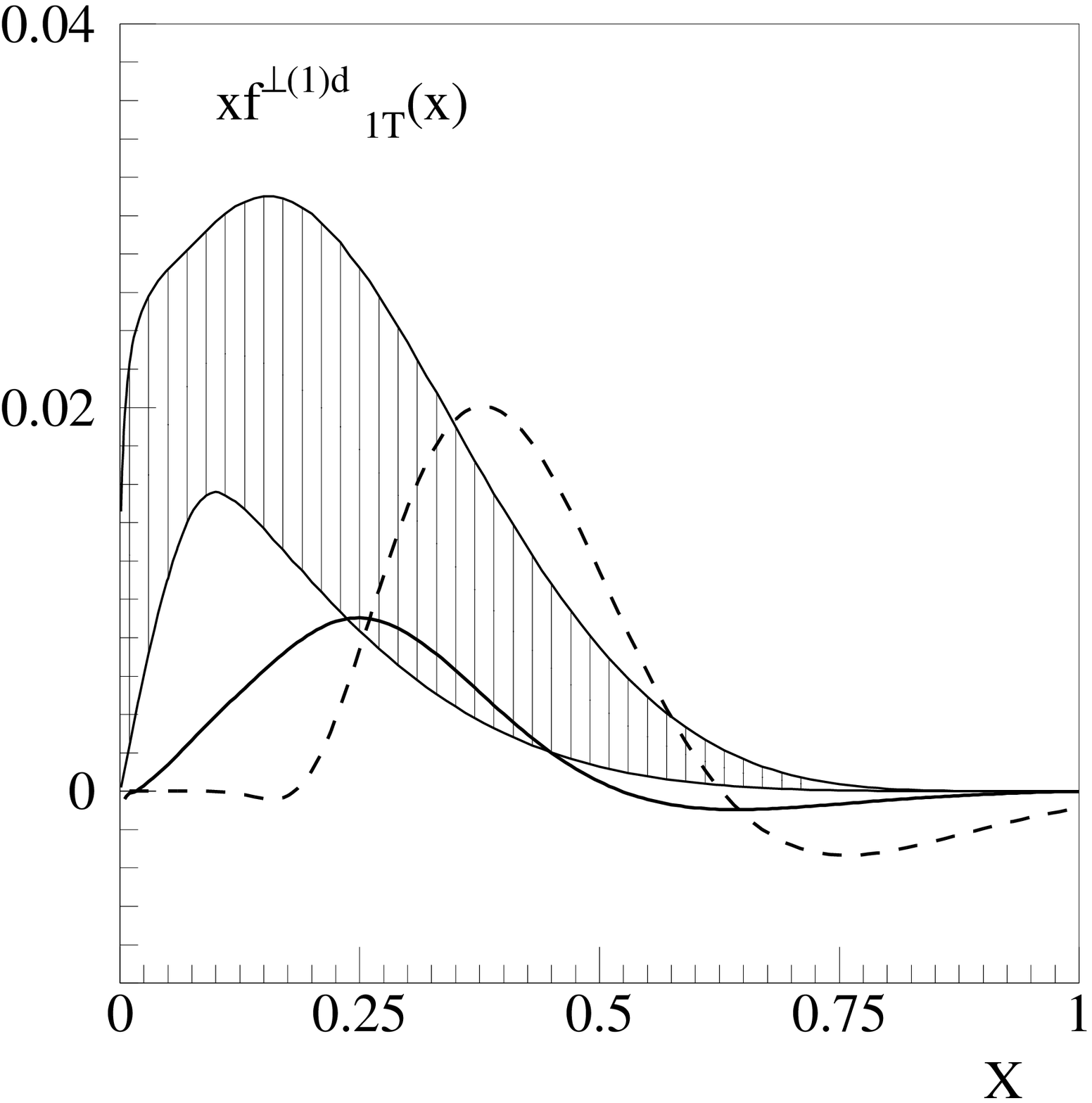}
\caption{
Left (right): the quantity $f_{1T}^{\perp (1) u(d) }(x) $, Eq. (\ref{momf}).
Dashed curve: IK at  $\mu_0^2$.
Full curve:  the evolved distribution at NLO.
Patterned area: parameterization by\cite{coll3} (see text).
}
\end{figure}
To evaluate numerically Eq. (\ref{start2}), $g$ (i.e.
$\alpha_s(Q^2)$) has to be fixed.
The prescription\cite{trvv} 
is used to fix $\mu_0^2$, according to the
amount of momentum carried by the valence quarks in the model.
Here, assuming that all the gluons and sea pairs in the proton
are produced perturbatively according to NLO evolution equations,
in order to have $\simeq 55 \% $ of the momentum
carried by the valence quarks at a scale of 0.34 GeV$^2$ 
one finds
that $\mu_0^2 \simeq 0.1$ GeV$^2$
if $\Lambda_{QCD}^{NLO} \simeq 0.24$ GeV.
This yields $\alpha_s(\mu_0^2)/(4 \pi) \simeq 0.13$ \cite{trvv}.
The results of the present approach
for the first  moments of the Sivers function, defined as
\begin{equation}
f_{1T}^{\perp (1) {\cal Q} } (x)
= \int {d^2 \vec k_T}  { k_T^2 \over 2 M^2}
f_{1T}^{\perp {\cal Q}} (x, {k_T} )~,
\label{momf}
\end{equation}
are given
by the dashed curves in Fig. 2.
They are compared
with a parameterization of the HERMES data,
taken at $Q^2=2.5$ GeV$^2$ :
The patterned area represents the $1-\sigma$ range
of the best fit proposed in Ref. \cite{coll3}.
The magnitude of the results
is close to that of the data,
although
they have a different shape: the maximum (minimum)
is predicted at larger values of $x$.
Actually $\mu_0^2$ is much lower, $Q^2 =2.5$ GeV$^2$. 
A proper comparison requires QCD evolution of TMDPDs,
 what is, to large extent, unknown.
We nevertheless perform a NLO evolution of the model
results assuming, for $f_{1T}^{\perp (1) {\cal Q} } (x)$,
the same anomalous dimensions of the unpolarized PDFs.
From the final result (full curve in Fig. 2),  one can see
that the  agreement with data
improves dramatically and the
trend is reasonably reproduced at least for $x \ge 0.2$.
Although the performed evolution is not exact, 
the procedure highlights the necessity 
of evolving the model results 
to the experiment scale and it suggests 
that the present results could be
consistent with data,  
still affected by large errors.


Properties of the Sivers function can be inferred from general principles. 
The Burkardt Sum Rule (BSR)
\cite{Burkardt:2004ur} states that, 
for a proton  polarized in the positive $y$ direction,
$\sum_{{\cal Q}=u,d} \langle k_x^{\cal{Q}} \rangle = 0$
with
\begin{equation}
\langle k_x^{\cal{Q}} \rangle = - \int_0^1 d x \int d \vec k_T
{k_x^2 \over M}  f_{1T}^{\perp \cal{Q}} (x, {k_T} )~,
\label{burs}
\end{equation}
and must be satisfied at any scale.
Within our scheme, at the scale of the model, it is found
$\langle k_x^{u} \rangle = 10.85$ MeV,
$\langle k_x^{d} \rangle = - 11.25$ MeV and, 
in order to have an estimate
of the quality of the agreement of our results with
the sum rule, we define the ratio
$r= 
| \langle k_x^{d} \rangle+
\langle k_x^{u}\rangle | /
| \langle k_x^{d} \rangle-
\langle k_x^{u} \rangle | $
obtaining $r \simeq 0.02$, so that we can say that our calculation
fulfills the BSR to a precision of a few percent.
One should notice that the agreement which is found
is better than that found in other model calculations
\cite{bacch,yuan},
especially for what concerns the fulfillment of the
Burkardt Sum Rule.

\section{The Sivers function from neutron
($^3$He) targets}

As explained in the previous section,
the experimental scenario which arises from the analysis
of SIDIS off transversely polarized
proton and deuteron targets \cite{hermes,compass} is 
puzzling. The data show
an unexpected flavor dependence in the azimuthal
distribution of the produced pions. 
With the aim at extracting the neutron information
to shed some light on the problem,
a measurement of SIDIS
off transversely polarized $^3$He has been addressed \cite{bro},
and two experiments, aimed at measuring the azimuthal asymmetries
in the production of leading $\pi^\pm$  from transversely
polarized $^3$He, are forth-coming at JLab \cite{ceb}.
Here, a realistic analysis of SIDIS
off transversely polarized $^3$He \cite{mio} is described.
The expressions
of the
Collins and Sivers contributions to the azimuthal
Single Spin Asymmetry (SSA) for the production
of leading pions have been derived, in impulse approximation (IA),
including the initial transverse momentum of the struck quark.
The final equations are involved and they are not
reported here. They can be found in \cite{mio}.
The same quantities have been then evaluated
in the kinematics of the  
JLab experiments.
Wave functions \cite{pisa} obtained within
the AV18 interaction \cite{av18} have been used for a realistic
description of the nuclear dynamics,
using overlap integrals evaluated in Ref. \cite{over},
and the nucleon structure has been described
by parameterizations of data or model calculations
\cite{ans,model}.
The crucial issue of extracting
the neutron information from $^3$He data
will be now discussed. 
As a matter of facts,
a model independent procedure, based
on the realistic evaluation
of the proton and neutron effective polarizations in $^3$He
\cite{old}, called respectively $p_p$ and $p_n$ in the following,
is widely used in  
DIS to take into account effectively  
the momentum and energy distributions
of the bound nucleons in $^3$He.
It is found that the same extraction technique
can be applied also in the 
kinematics of the proposed experiments, although
fragmentation functions, not only parton
distributions, are involved, as it can be seen
in Figs. 1 and 2. In these figures,
the free neutron asymmetry used as a model in the
calculation, given by a full line, is compared with two other quantities.
One is:
\begin{equation}
\bar A^i_n \simeq {1 \over d_n} A^{exp,i}_3~,
\label{extr-1}
\end{equation}
where $i$ stands for ``Collins'' or ``Sivers'',
$A^{exp,i}_3$ is the result of the full calculation, 
simulating data, and $d_n$ is the neutron dilution
factor. The latter quantity is defined as follows, for a neutron $n$
(proton $p$) in $^3$He:
\begin{eqnarray}
d_{n(p)}(x,z)=
{\sum_q e_q^2
f^{q,{n(p)}} 
\left ( x \right )
D^{q,h} \left ( z  \right )
\over
\sum_{N=p,n}
\sum_q e_q^2
f^{q,N} 
( x )
D^{q,h} 
\left ( z \right )
}
\label{dilut}
\end{eqnarray}
and, depending on the standard parton
distributions, $ f^{q,N} ( x )$,
and fragmentation functions, $D^{q,h} 
\left ( z \right )$,
is experimentally known (see \cite{mio} for details). 
$\bar A^i_n $ is given by the dotted curve in the figures.
The third curve, the dashed one, is given by 
\begin{equation}
A^i_n \simeq {1 \over p_n d_n} \left ( A^{exp,i}_3 - 2 p_p d_p
A^{exp,i}_p \right )~,
\label{extr}
\end{equation}
i.e. $^3$He is treated as a nucleus
where the effects of its 
spin structure, of
Fermi motion and binding, can be taken care
of by parameterizing 
$p_p$ and $p_n$.
\begin{figure}
\includegraphics[width=.49\textwidth]{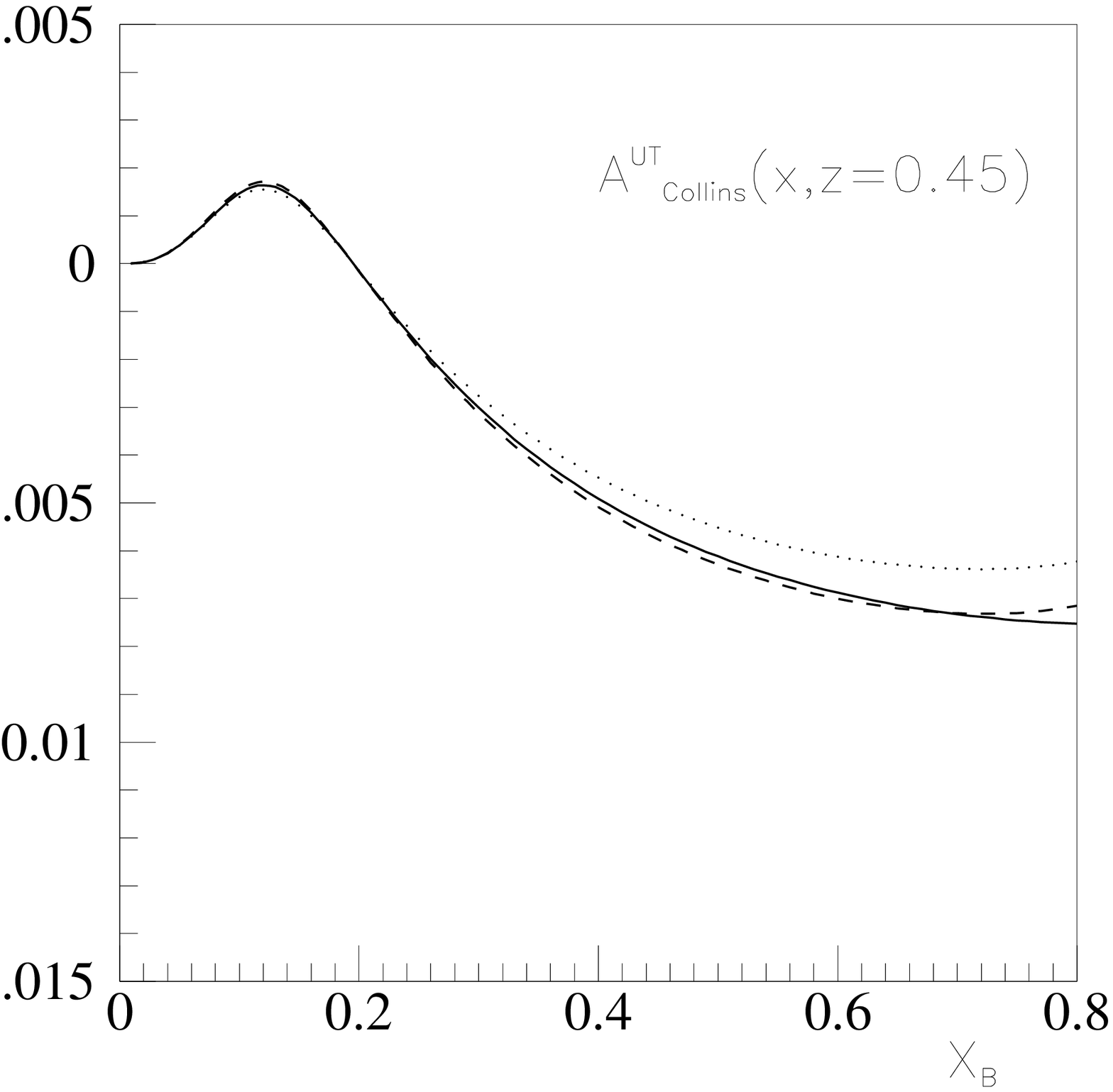}
\includegraphics[width=.49\textwidth]{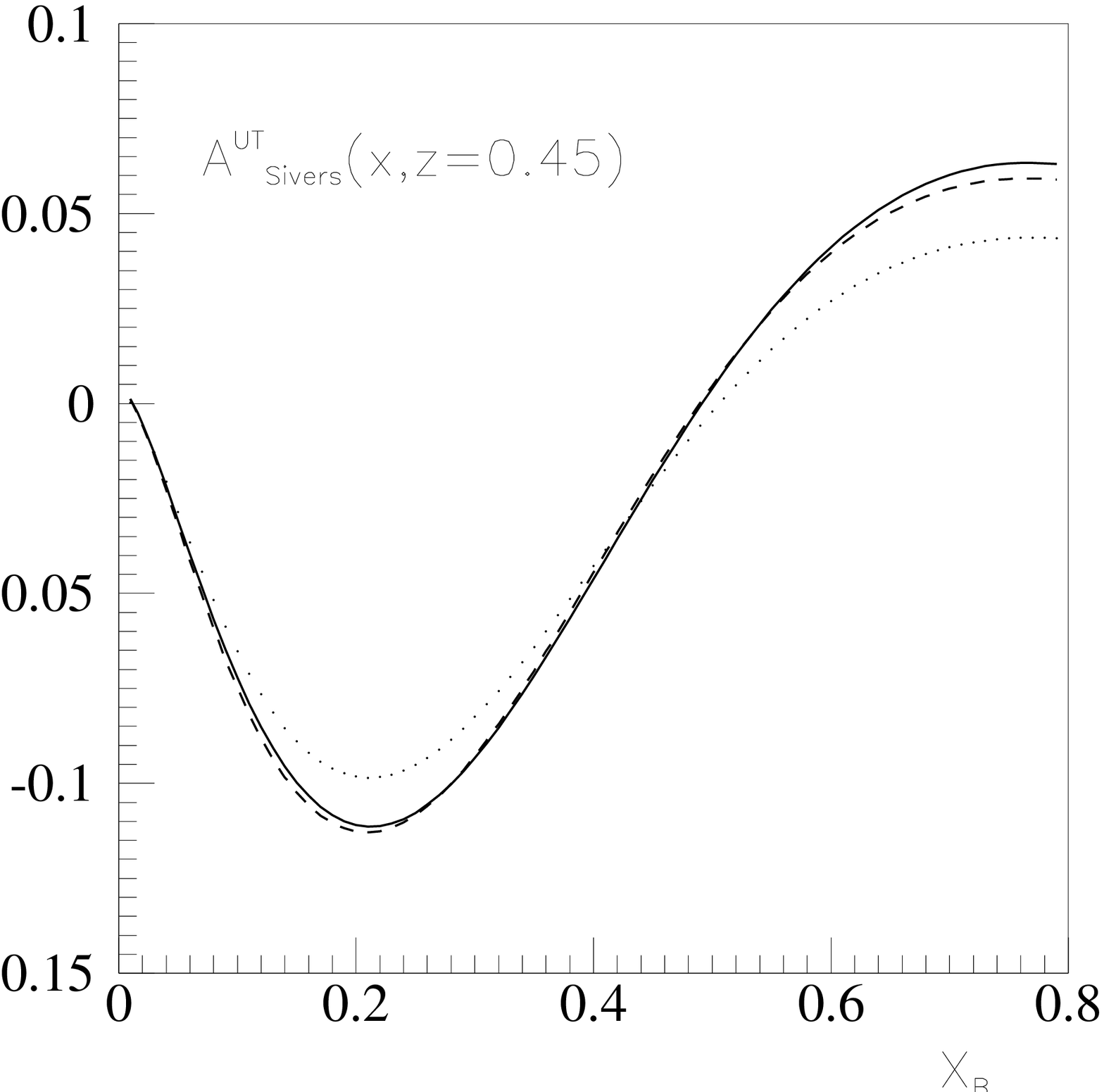}
\caption{Left (right)
The model neutron Collins (Sivers) asymmetry for
$\pi^-$ production
(full) in JLab kinematics, and the one extracted
from the full calculation taking into account
the $p_p$
(dashed), or neglecting it (dotted). 
The results are shown for {$z$}=0.45 and
$Q^2= 2.2$ GeV$^2$, typical values
in the kinematics of the JLab experiments.}
\end{figure}
One should realize that
Eq. (\ref{extr-1}) is the relation which should hold
between the $^3$He and the neutron SSAs if there were no nuclear effects,
i.e. the $^3$He nucleus were a system of free nucleons in a pure $S$ wave.
In fact, Eq. (\ref{extr-1}) can  be obtained from Eq. (\ref{extr}) by 
imposing $p_n=1$ and $p_p=0$.
It is clear from the figures that the difference 
between the full and dotted curves,
showing the amount of nuclear effects, is sizable,
being around 10 - 15 \% for any experimentally relevant $x$ and $z$,
while the difference between the dashed
and full curves reduces drastically
to a few percent, showing that the extraction
scheme Eq. (\ref{extr}) takes safely into account
the spin structure of $^3$He, Fermi
motion and binding effects. 
This important result is due to the kinematics
of the JLab experiments, which helps in two ways.
First of all, to favor pions from current fragmentation, 
$z$ has been chosen in the range $0.45 \leq z \leq 0.6$,
which means that only high-energy pions are observed.
Secondly, the pions are detected in a narrow cone around the direction
of the momentum transfer. As it is explained in \cite{mio},
this makes nuclear effects in the fragmentation 
functions rather small. The leading nuclear effects are then 
the ones affecting the parton distributions, already found
in DIS, and can be taken into account
in the usual way, i.e., using Eq. (\ref{extr}) for the extraction of the
neutron information. In the figures,
one should not take the 
shape and size of the asymmetries seriously,
being the obtained quantities 
strongly dependent on the models chosen for the unknown distributions
\cite{model}.
One should instead consider the difference between
the curves, a model independent
feature which is the most relevant outcome of the present
investigation. 
Eq. (\ref{extr}) is therefore a valuable tool
for the experiments \cite{ceb}.
The evaluation of 
final state interactions effects and the inclusion
of more realistic models of the nucleon structure are in progress.


\section*{Acknowledgments}
This work is supported in part by the INFN-CICYT agreement,
by the Generalitat Valenciana under the contract
AINV06/118; by the Sixth Framework Program of the
European Commission under the Contract No. 506078 (I3 Hadron Physics);
by the MEC (Spain) (FPA 2007-65748-C02-0, AP2005-5331 and PR2007-0048).


\bibliographystyle{ws-procs9x6}
\bibliography{ws-pro-sample}

\end{document}